# **Title:** Interlayer Exciton Diode and Transistor

***Author Names:*** *Daniel N. Shanks[1], Fateme Mahdikhanysarvejahany[1], Trevor G. Stanfill[1], Michael R. Koehler[2], David G. Mandrus[3-5], Takashi Taniguchi[6], Kenji Watanabe[7], Brian J. LeRoy[1], and John R. Schaibley[1]\**

**Affiliations:**

[1]Department of Physics, University of Arizona, Tucson, Arizona 85721, USA

[2]IAMM Diffraction Facility, Institute for Advanced Materials and Manufacturing, University of Tennessee, Knoxville, Tennessee 37920, USA

[3]Department of Materials Science and Engineering, University of Tennessee, Knoxville, Tennessee 37996, USA

[4]Materials Science and Technology Division, Oak Ridge National Laboratory, Oak Ridge, Tennessee 37831, USA

[5]Department of Physics and Astronomy, University of Tennessee, Knoxville, Tennessee 37996, USA

[6]International Center for Materials Nanoarchitectonics, National Institute for Materials Science, 1-1 Namiki, Tsukuba 305-0044, Japan

[7]Research Center for Functional Materials, National Institute for Materials Science, 1-1 Namiki, Tsukuba 305-0044, Japan

**\*Corresponding Author:** John Schaibley, johnschaibley@email.arizona.edu

**Abstract:**

Controlling the flow of charge neutral interlayer exciton (IX) quasiparticles can potentially lead to low loss excitonic circuits. Here, we report unidirectional transport of IXs along nanoscale electrostatically defined channels in an MoSe₂-WSe₂ heterostructure. These results are enabled by a lithographically defined triangular etch in a graphene gate to create a potential energy "slide". By performing spatially and temporally resolved photoluminescence measurements, we measure smoothly varying IX energy along the structure and high-speed exciton flow with a drift velocity up to $2 \times 10^6$ cm/s, an order of magnitude larger than previous experiments. Furthermore, exciton flow can be controlled by saturating exciton population in the channel using a second laser pulse, demonstrating an optically gated excitonic transistor. Our work paves the way towards low loss excitonic circuits, the study of bosonic transport in one-dimensional channels, and custom potential energy landscapes for excitons in van der Waals heterostructures.

**Main Text:**

Excitonic circuits are optoelectronic devices where excitons are used to encode and process information in a manner analogous to electrons in electronic circuits[1]. Two basic building blocks for excitonic circuits are diodes, which must demonstrate unidirectional flow of excitons[2–4], and transistors, which must demonstrate controlled gating of excitons[5–8]. Excitons comprise a charge negative electron bound to a charge positive hole resulting in a charge neutral quasiparticle with integer spin. The charge neutrality of the exciton makes it insensitive to long range Coulomb scattering mechanisms, offering the opportunity to realize transport with negligible Ohmic losses[5,9,10]. The >100 meV binding energy of excitons in van der Waals heterostructures shows potential for room temperature operation of such devices[11], previously unattainable in similar coupled quantum well systems.

In this work, we demonstrate both a high speed excitonic diode and an excitonic transistor based on interlayer excitons (IXs) in transition metal dichalcogenide (TMD) heterostructures. We utilize the IX enabled by the type-II band alignment between monolayer $WSe_2$ and $MoSe_2$ which has been shown to exhibit outstanding optoelectronic properties[12–16]. Previous works have shown that the IX exhibits a long (> 1 ns) lifetime, as well as a highly tunable energy via electrostatic potentials, arising from the spatial separation between the electron and hole layers[12,17–19]. Excitons on TMDs also have the ability to be valley polarized to realize valleytronic devices[20–23]. The ability to weakly control exciton flow by micron-sized backgates has been previously shown[11,19,24–26], but nanoscale control of exciton flow has not been achieved. Here, we show that by using a nanopatterned graphene topgate, we can induce sharply varying electric fields to create quasi-one dimensional channels for high speed, controlled exciton flow with a single gate connection[27]. This IX flow is driven by the electric field-dipole moment potential energy and IX-IX repulsion[12,19,28]. Further, we demonstrate a excitonic gating operation, where a density of "control" excitons is used to gate the flow of "signal" excitons, demonstrating a pure IX transistor[7,8]. The demonstrated architecture opens the door for the creation of custom and complex potential energy landscapes for IXs in van der Waals heterostructures using nanopatterned graphene gates[29–32].

In order to increase the mobility of IXs, we suppress the moiré potential by placing a hexagonal boron nitride (hBN) spacer between the TMD layers[28,33–36]. This increases the IX energy from 1.33 to 1.41 eV (Fig. S1), and increases the IX dipole moment to 1.1 nm (Fig. S2), in excellent agreement with the 0.5 nm bilayer hBN separator layer plus the 0.6 nm dipole moment of the IX without the hBN separator[19,24,36], and allowing for greater tunability of exciton energy via the electric field-dipole moment potential energy. The schematic of the excitonic device is shown in Fig. 1a-b, with an optical microscope image of the device in Fig. S3. The TMD heterostructure is encapsulated by ~35 nm hBN to provide a clean, flat substrate, and few layer graphene (FLG) gates to control the electric field.

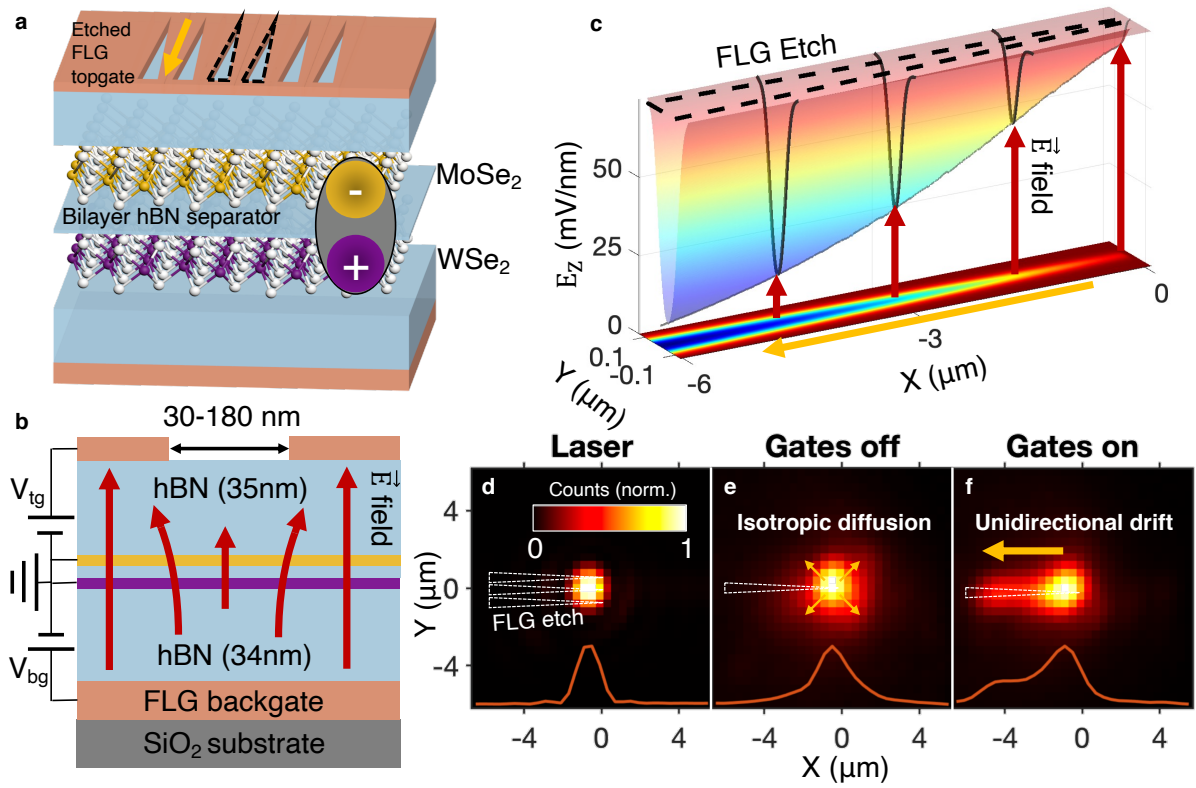

**Figure 1.** Schematic for IX diode and transistor device. (a-b) Device schematic, with contacts attached to FLG gates, and triangular "slide" structures etched into FLG topgate to apply spatially varying electric field. (c) COMSOL calculation of the vertical electric field component under the etched graphene, at the heterostructure layer height. $V_{tg}$ = -7 V, $V_{bg}$ = 0 V. Black dashed line indicates profile of FLG etch, black solid curves show linecuts along y = 0 μm, x = -1.5, -3.0, -4.5 μm. Red lines in (b) and (c) depict the $\vec{E}$ field lines of the spatially varying electric field. (d) CCD image of excitation laser intensity profile, centered at x = -0.5 μm. White triangles indicate location of FLG etch. (e-f) Spatially resolved IX PL signal with gates off ($V_{tg}$ = $V_{bg}$ = 0 V) (e) and gates on ($V_{tg}$ = -7 V, $V_{bg}$ = +3.5 V) (f). Orange lines show centerline cut of PL signal at y = 0 μm. Sharp tip of triangular FLG etch is located at x = 0 μm in (c-f). Gold Arrows in (a, c, f) indicate direction of induced IX current, gold arrows in (e) indicate isotropic diffusion of IXs.

After the heterostructure is constructed, the top graphene gate is nano-patterned by high resolution electron beam lithography and reactive ion etching[27]. The topgate graphene etch consists of long, skinny, isosceles triangles, with a base width of 175 nm, and height of 6 μm, spaced 220 nm apart (Fig. S3). When a top gate voltage is applied to the sample, the patterned FLG creates a spatially varying electric field, which generates a smooth potential energy for profile for IXs[11,24,27]. The electric field dip under the triangles goes from shallow on the narrow end to deep on the wider end, which is confirmed by confocal spatial photoluminescence (PL) gate maps (see Fig. S4). This creates a potential energy landscape for IXs like a playground slide, shown in Fig. 1c, where IXs can be created by laser excitation at the top of the slide, and flow along the 1D channels to the base of the triangle.

We performed spatially resolved PL spectroscopy to measure the unidirectional flow of excitons. Figure 1d shows a spatial map of the excitation laser position, placed near the sharp tip of the triangular etch. All PL measurements were performed at 6 K, with a 225 μW, 670 nm continuous wave diode laser unless otherwise noted. Figures 1e-f show the spatial distribution of the IX PL intensity with the induced slide potential off (e) and on (f). In Fig. 1e, we used an 800-900 nm band pass (800-850 nm in Fig. 1f) filter to block laser light, monolayer exciton emission, and emission from the hBN-non separated heterostructure region, and pass the hBN-separated IX PL. When both graphene gates are grounded, we see isotropic diffusion of IXs away from the laser spot, driven primarily by IX-IX repulsion[28]. In Fig 1f, the electric field was applied in the opposite direction of the IX dipole moment, and the IX energy is increased to 1.45-1.55 eV depending on location within the slide. We observe highly anisotropic PL, showing unidirectional IX current from the region of high electric field to low, with similar data from another etched graphene structure shown in Fig. S5. For these applied gate voltages, the strength of the electric field varies from 95 mV/nm at its strongest point above the top of the slide to 40 mV/nm at the bottom. We apply a 3.5 V backgate in this scan in order to shift the lowest IX energy above 850 nm, in the range of the bandpass filter. We find that the applied backgate voltage uniformly shifts the IX energy across the entirety of the slide structure, but does not significantly change the spatial dependence of the PL.

We directly measured the position dependence of the IX energy using spatial and energy resolved PL. Here, we align the long direction of the triangular etch parallel to the lines of our diffraction grating, allowing one axis of the CCD to measure position and the other axis to measure photon energy. Figure 2a shows this PL measurement when IXs are excited at x = -1 µm below the top of the slide. Overlaid on this data, the blue curve in Fig. 2a shows the calculated IX energy using the measured zero field energy, plus the measured dipole moment multiplied by the COMSOL calculated electric field along the center of the slide, y = 0 µm in Fig. 1c. The COMSOL model agrees very well with both the measured PL energy and position. This agreement between theoretical and experimental results demonstrates the accuracy of our architecture and opens the door for custom 2D potential energy landscapes using nanopatterned graphene gates to spatially control IX energy.

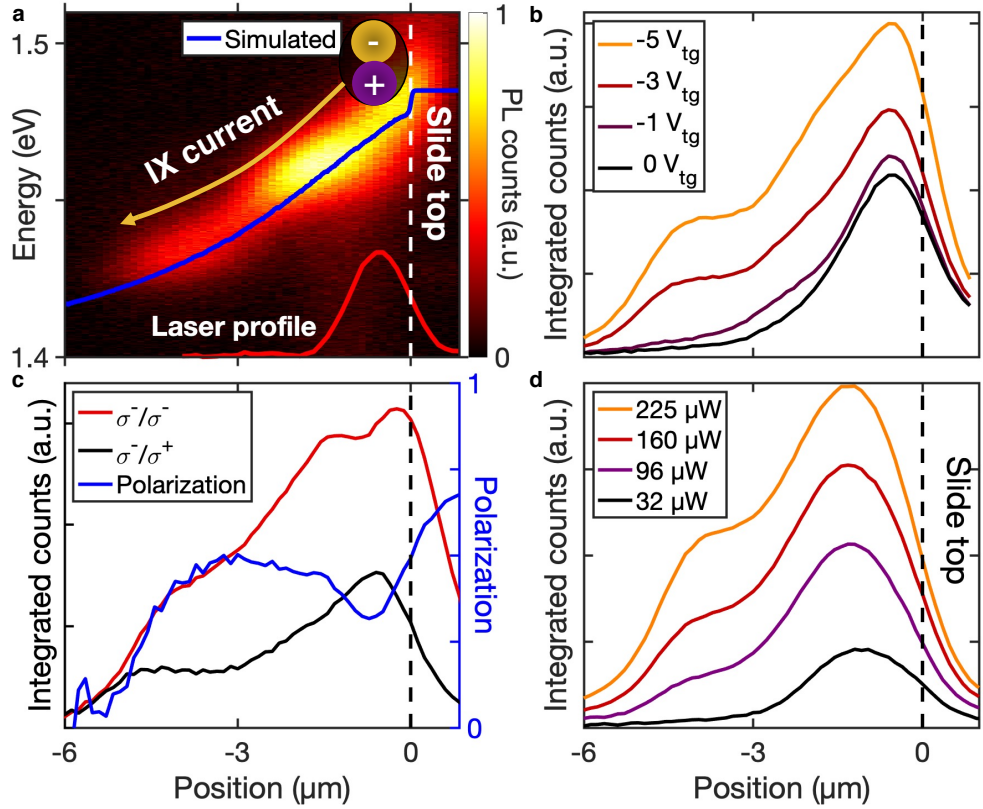

**Figure 2.** Spatially resolved PL of IXs. (a) Spatial calculation (blue line) of IX energy overlaid on PL emission energy as a function of position. Laser excitation profile (red line) is positioned near the top of the slide, x = -1 μm, $P_{exc}$ = 112 μW. The top of the slide corresponds to position x = 0 μm, indicated by white and black dashed lines. (b) Spectrally integrated counts of PL emission over the IX energy range (1.4 to 1.51 eV) as the IX potential is changed by gate voltage, $V_{bg}$ = 0 V. At $V_{tg}$ = 0 V, there is no induced potential, and isotropic diffusion as expected. (c) Co- (red) and cross-(black) circularly polarized PL spectra as a function of position, integrated over emission energy with the slide potential turned on. Blue line shows polarization, calculated as $\frac{\sigma^- - \sigma^+}{\sigma^- + \sigma^+}$ for $\sigma^-$ laser excitation, using 720 nm excitation at 60 μW. (d) Spatial PL emission integrated over emission energy for four different laser excitation powers. $V_{tg}$ = -7 V, $V_{bg}$ = 0 V in (a), (c-d), laser excitation spot x = -1 μm in (a-c), -1.5 μm in (d).

We confirmed that the unidirectional flow of excitons is induced by the electric field by taking spatial PL measurements with varying gate voltage. Figure 2b plots this position dependent PL, showing increased IX current along the slide with increasing field. Figure S6 shows spatial and energy resolved PL, showing that the slope of the IX potential energy curve increases proportionally with topgate voltage. In Fig. 2c, we show polarization resolved PL along the slide to test the feasibility of this architecture for valleytronic applications, pumping with a 1.72 eV laser. The IX PL is primarily co-circularly polarized with the laser excitation (see Fig. S1), as seen in previous studies on hBN-separated $WSe_2$-$MoSe_2$ heterostructures[24,36]. We note that we observe nearly constant PL polarization of ~50%, showing that the valley polarization remains intact during induced exciton motion. Figure 2d shows the spatial dependence of the IX PL with varying pump powers, which shows that high power (~100 μW) resulting in high IX density and high IX-IX repulsion force is required to realize long range IX current. Figures S6 and S7 show spatial and energy resolved PL data over which Fig. 2b-d are integrated. Figure S8 shows the temperature dependence of the PL signal, showing no characteristic change of IX diffusion up to 35 K. In order to study effects from spatially varying doping under the etched graphene, we performed doping dependent confocal photoluminescence on an unpatterned area of the heterostructure, shown in Fig. S9. We observe that for the range of gate voltages applied in Fig. 2, doping of the sample does not significantly affect IX energy.

We verified the unidirectional flow of IXs by performing PL measurements with various excitation positions along the slide. Figure 3 shows spatially and spectrally resolved PL for three different excitation spots. In Fig. 3a, the laser position was located at x = -2 μm below the top of the diode, and we primarily see exciton flow to the left, to lower electrical potential energy. We see a small amount of exciton flow towards the top of the slide due to IX-IX repulsion; however, the PL signal to the right of the laser spot was less than one third of the total PL signal, indicating that two thirds of the IXs move down the diode. As the excitation spot was moved in Figs. 3b-c, we continued to observe anisotropic diffusion of excitons, with preferential flow to lower potential energy. With the excitation position at the bottom edge of the diode in Fig. 3c, there is no IX diffusion beyond $x = -2$ µm, confirming that even for high power, IX-IX repulsion cannot drive IX flow back up the device.

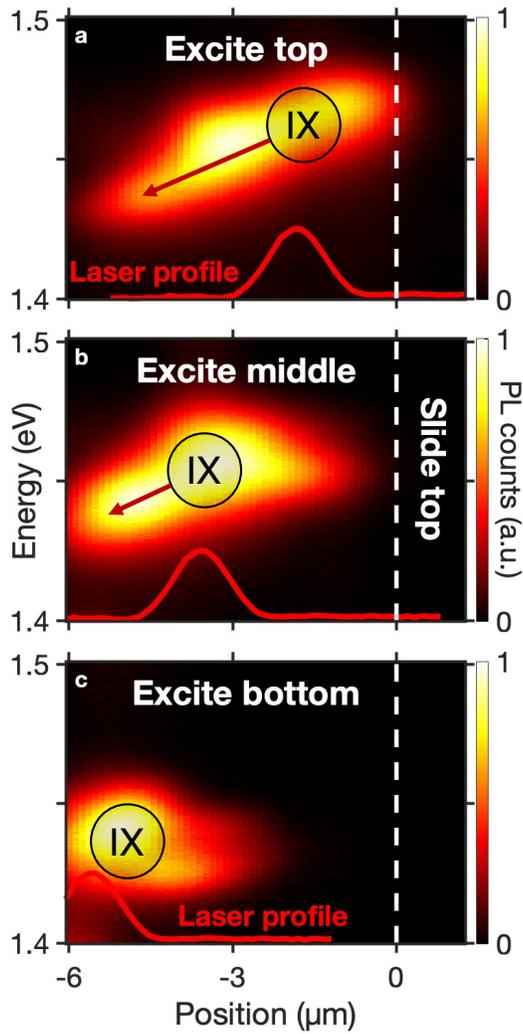

**Figure 3.** Excitonic diode. Spatial and energy resolved PL for three laser positions, showing diode-like, unidirectional drift of IXs. Laser excitation position near x = -2 μm (a), -4 μm (b), -6 μm (c), with the top of the slide at x = 0 μm, indicated by the white dashed line. $V_{tg}$ = -7 V, $V_{bg}$ = 0 V. Grey circles indicate initial IX position, and red arrows indicate IX current. Red gaussian profile shows excitation laser position for each scan. PL emission ends at x = -6 μm, corresponding to the physical edge of the heterostructure.

We used time-resolved PL (time correlated single photon counting) to measure the IX transport time across the device. Here, we detected PL from the collection spot at the bottom of the slide and used a pulsed laser to inject IXs at varying locations along the structure. We filtered the IX PL both spatially and spectrally, using a confocal pinhole and the exit slit of the spectrometer to collect only PL emission with the wavelength corresponding the collection spot. Figure 4a shows the rise of this time-resolved PL. We calculate an exciton drift velocity of $1.6 \times 10^5$ cm/s in this data by dividing the $\Delta x$ between laser excitation positions by the $\Delta t$ delay in exciton emission from the collection spot, determined by the time at which the PL counts reach half of their maximum. We find that this exciton drift velocity depends both on applied gate voltage and laser power, indicating that both the induced electrical potential energy and IX-IX repulsion impact exciton drift velocity. For the highest gate voltage and excitation power used in this experiment, the IX speed is measured at $2 \times 10^6$ cm/s (Fig. S10), an order of magnitude larger than values reported in *Sun et al.*[28] involving IX drift driven only by 2D IX-IX repulsion, without any confining potential. We attribute the increased drift velocity to two sources. The first is the slide IX potential induced by the spatially varying electric field. The second is the strong confinement of the IXs to narrow, quasi-one dimensional channels which increases the repulsive IX-IX interaction strength due to the increased IX density. Figure S4 shows evidence that nearly all IXs in the vicinity of the slide structure, which start with homogenous density, fall into the 1D channels, creating significantly higher IX density within the slides compared to unconfined IX diffusion studied in *Sun et al.*[28]. Figure 2d shows that excitation laser intensity significantly affects IX diffusion, further supporting the conclusion that IX density plays a significant role in IX velocity along the slide.

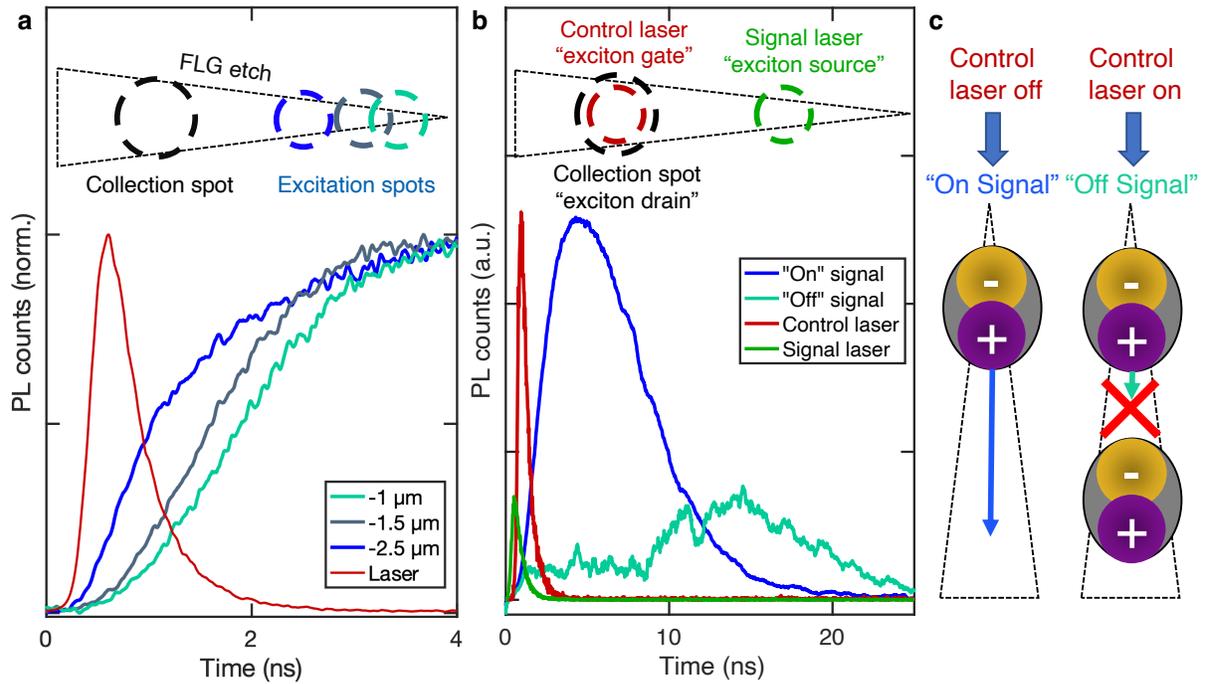

**Figure 4.** IX drift velocity measurement and transistor operation. (a) Time resolved PL, with collection from the bottom of the slide, near x = -4 μm. Excitation positions x = -1, -1.5, -2.5 μm represent the distance from laser excitation position to the top of the slide located at x = 0 μm. Excitation laser wavelength is 720 nm, at 60 μW. (b) Depiction and data of the optically gated exciton transistor. Green, red and black dashed circles show physical locations of the signal laser (x = -2 μm), control laser, and PL collection spot (x = -4 μm), relative to the triangular FLG etch. Green and red lines show time delay and relative laser power (not to scale) of signal and control laser pulses. Blue (cyan) line shows PL from IXs created by signal laser with the control laser off (on). $V_{tg}$ = -3 V, $V_{bg}$ = 0 V in a and b. (c) Depiction of IX flow, determined by control laser injecting IX population at the bottom of the slide.

We also demonstrate that IX current can be gated by the presence of IXs, realizing an optically gated excitonic transistor[7,8]. The inset of Fig. 4b depicts the two-pulse experiment, where the first "signal" laser pulse creates IXs at the top of the device, which flow to the bottom, unless a second "control" laser pulse is turned on, creating IX density at the bottom of the device which blocks the flow of signal excitons, shown by cartoon in Fig. 4c. The blue curve in Fig. 4b shows confocal time resolved PL from the bottom of the transistor, from only the low power (10 μW) signal laser pulse near the top of the transistor, creating an "on" IX signal, where IXs freely flow down the slide. Alternatively, we can turn on a higher power (55 μW) control laser pulse at the collection spot, filling the exciton states within the bottom of the device (see Fig. S11). We treat the control-only PL signal as a background, which we subtract from the signal when both lasers are on, and measure this as the "Off" signal (cyan curve in Fig. 4b). This IX gated flow of IXs demonstrates the operation of an exciton gated IX transistor, a basic building block of excitonic circuits. We report a maximum integrated On/Off ratio of 8:1 for t = 0 to 10 ns, which varies depending on signal laser intensity and gate voltage. The pulse width of the laser used in this experiment was <70 ps, resulting in an effective switching rate of ~14 GHz.

In summary, we have introduced a novel architecture to control the potential energy landscape and flow of IXs in TMD heterostructures. We demonstrated a highly directional IX diode, and an excitonic transistor using IX-IX repulsion to control the flow of excitons. While previous methods to drive anisotropic exciton transport in 2D semiconductors using strain engineering[9], or surface acoustic waves[10] have shown promising results, our approach using nanopatterned graphene creates to smallest physical channel for excitons, shows the highest measured IX speed, and is the first to show optically gated exciton transport. Our experimental results are in excellent quantitative agreement with our theoretical modelling showing that the electric fields generated by our nano-patterned graphene gates are accurate and reliable. The novel excitonic architecture developed in this work has broad application to the realization of low power consumption IX circuits[1] at room temperature[11] that use spin or valley degrees of freedom to encode and process information[22]. The application of nanopatterned graphene provides a new path to perform fundamental studies of bosonic transport in customizable potential energy profiles, and properties of 2D materials in nanoscale spatially modulated electric fields.

**Methods:**

**Device fabrication**: 2D layers in the device were mechanically exfoliated from bulk crystals onto 285 nm or 90 nm $SiO_2$/Si wafers. They were then identified by optical microscopy and characterized for thickness and cleanliness with atomic force microscopy (AFM). TMD layers were angle aligned, with their crystal axes determined by second harmonic generation measurements[37,38]. Layers were transferred on top of one another using a dry transfer technique using a polycarbonate (PC) film on polydimethylsiloxane (PDMS)[39], aligned to one another under an optical microscope, and placed on a chip with a pre-patterned gold grid for electron beam lithography (EBL) alignment. The device structure from top to bottom is etched few layer graphene (FLG)/35 nm hBN/1L $MoSe_2$/2L hBN/1L $WSe_2$/34 nm hBN/FLG.

The FLG top layer was nano-patterned via EBL in a 100 kV acceleration Elionix Model ELS-7000 system, and reactive ion etching in a Plasmatherm Versaline DSE III. A thin layer of 950 poly(methylmethacrylate) (PMMA) A2 was spun at a rate of 3k rpm to produce an 80 nm thick film on top of the heterostructure. EBL was performed with an aperture width of 60 μm and electron beam current of 50 pA. and PMMA was subsequently developed in a 1:4 MIBK:IPA solution for 30 seconds, with light stirring. We then use an inductively coupled plasma reactive ion etch to etch the graphene, with 50 sccm $O_2$ gas at a pressure at 100 mTorr and 100 W AC power for 5 seconds. The PMMA was washed away in acetone and the sample was cleaned in isopropyl alcohol. We then use EBL again using a 30 kV FEI Inspect SEM to define contacts and thermally evaporate 10 nm Cr/50 nm Au to electrically connect to the graphene gates and TMD layers.

**Optical measurements**: All PL spectroscopy was performed in a Montana Instruments optical cryostat at 6 K. The sample was excited with a 670 nm diode laser except for the time and polarization resolved measurements which were performed with a CW tunable Ti:sapphire laser at 720 nm (M Squared), or a 78 MHz pulsed supercontinuum laser (NKT), with a 10 nm bandwidth wavelength filter set to 720 nm, and rep rate reduced to either 39 MHz (Fig. 4a) or 26 MHz (Fig. 4b) using a pulse picker. The pulse width of the laser is < 70 ps. Gate voltages were applied to the graphene top and bottom gates, while the TMD layers were connected to ground. The confocal PL measurements were performed with a 0.6 NA 40x objective and a confocal pinhole which results in a collection diameter of 1.5 µm on the sample. The steady state PL signal was measured with a grating-based spectrometer and cooled CCD camera. In the polarization resolved measurements, combinations of polarizers and achromatic wave plates were used. The time resolved measurements were performed via time resolved single photon counting with a silicon avalanche detector and picosecond event timer (Picoquant).

**COMSOL simulations**: COMSOL simulations used a 3D electrostatic model using known values for the in-plane and out-of-plane dielectric constants of the hBN and TMD layers, and hBN thicknesses that matched the AFM data of the individual 2D layers. Voltages applied to top and back gates were set at the graphene-hBN interfaces, while the top and bottom interfaces of the TMD heterostructure were grounded. Electric field data were calculated by averaging the field above and below the TMD heterostructure layer.

**Supporting Information**:

Device characterization, images, AFM topography, extended spatially and energetically resolved PL, time resolved PL and two-laser PL experiment off of the slide structure.

**Corresponding Author:**

Correspondence and requests for materials should be addressed to John Schaibley, johnschaibley@email.arizona.edu.

**Author Contributions:**

DNS, JRS, and BJL conceived the project. JRS and BJL supervised the project. DNS and FM fabricated the structures, assisted by TGS. DNS modelled the structures and performed the experiments. DNS analyzed the data with input from JRS and BJL. MRK and DGM provided and characterized the bulk $MoSe_2$ and $WSe_2$ crystals. TT and KW provided hBN crystals. DNS, JRS and BJL wrote the paper. All authors discussed the results.

**Funding Sources:**

JRS and BJL acknowledge support from the National Science Foundation Grant. Nos. ECCS-2054572 and DMR-2003583 and the Army Research Office under Grant no. W911NF-20-1-0215. JRS acknowledges support from Air Force Office Scientific Research Grant Nos. FA9550-18-1-0390 and FA9550-21-1-0219. BJL acknowledges support from the Army Research Office under Grant no. W911NF-18-1-0420. DGM acknowledges support from the Gordon and Betty Moore Foundation's EPiQS Initiative, Grant GBMF9069. K.W. and T.T. acknowledge support from JSPS KAKENHI (Grant Numbers 19H05790, 20H00354 and 21H05233). Plasma etching was performed using a Plasmatherm reactive ion etcher acquired through an NSF MRI grant, award no. ECCS-1725571.

**Data availability:**

All data needed to evaluate the conclusions in the paper are present in the paper or the Supporting information. Source or additional data related to this paper may be requested from the corresponding authors.

**Competing Interests:**

The authors declare no competing interests.

**Supporting Information**

# Interlayer Exciton Diode and Transistor

Daniel N. Shanks[1], Fateme Mahdikhanysarvejahany[1], Trevor G. Stanfill[1], Michael R. Koehler[2], David G. Mandrus[3-5], Takashi Taniguchi[6], Kenji Watanabe[7], Brian J. LeRoy[1], and John R. Schaibley[1]*

**Affiliations:**

[1]Department of Physics, University of Arizona, Tucson, Arizona 85721, USA

[2]IAMM Diffraction Facility, Institute for Advanced Materials and Manufacturing, University of Tennessee, Knoxville, Tennessee 37920, USA

[3]Department of Materials Science and Engineering, University of Tennessee, Knoxville, Tennessee 37996, USA

[4]Materials Science and Technology Division, Oak Ridge National Laboratory, Oak Ridge, Tennessee 37831, USA

[5]Department of Physics and Astronomy, University of Tennessee, Knoxville, Tennessee 37996, USA

[6]International Center for Materials Nanoarchitectonics, National Institute for Materials Science, 1-1 Namiki, Tsukuba 305-0044, Japan

[7]Research Center for Functional Materials, National Institute for Materials Science, 1-1 Namiki, Tsukuba 305-0044, Japan

**\*Corresponding Author:** John Schaibley, johnschaibley@email.arizona.edu

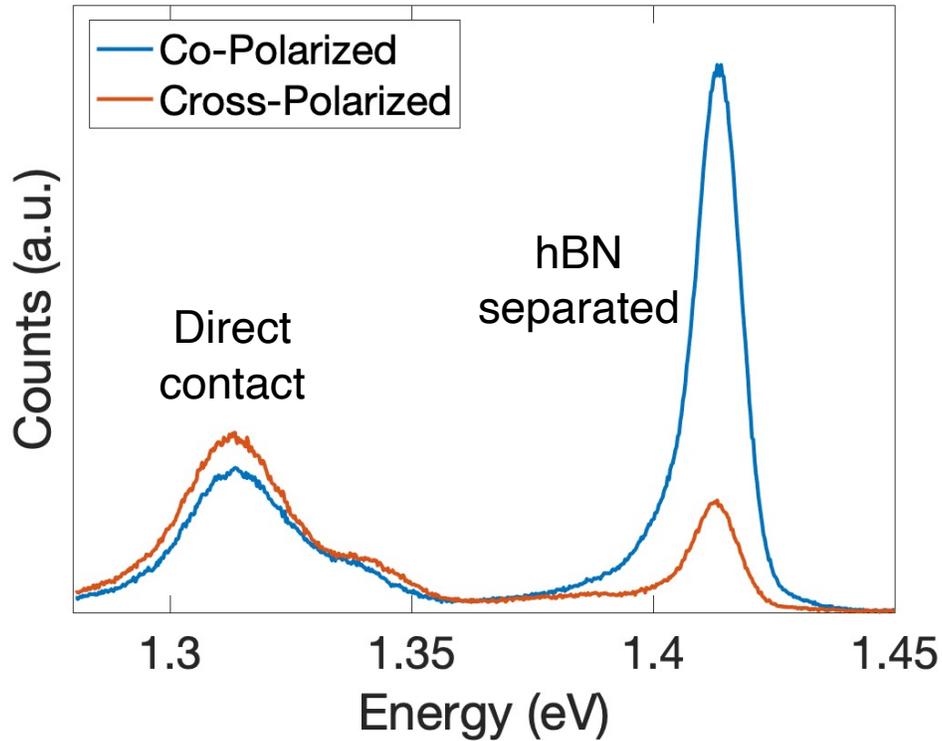

**Figure S1. Circular polarization resolved PL on and off the hBN separated TMD heterostructure.** Co (orange) and cross (blue) circularly polarized PL emission with left-hand circularly polarized light from both the hBN separated region and the region where the two TMD monolayers are in direct contact with one another. Excitation laser was 720 nm and an excitation power of 20 μW. The polarization on the direct contact region is primarily cross-circularly polarized, indicating that our sample is near 0-degree aligned (R-type).

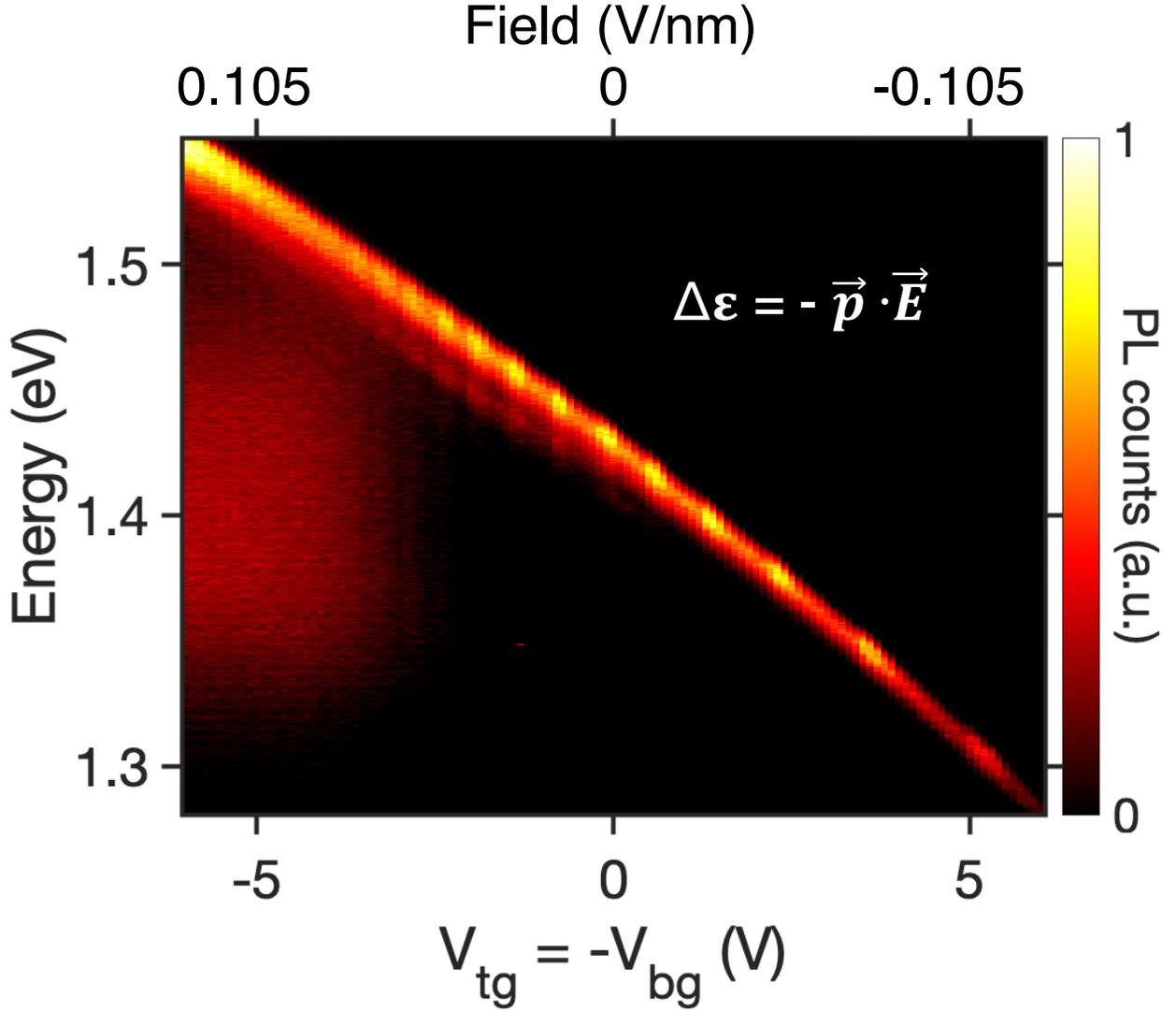

**Figure S2. Confocal PL gate map off the etched graphene area.** IX PL signal from the hBN separated region of the device, with topgate and backgate set to equal magnitude, opposite sign voltage. This applies an electric field without doping the heterostructure, given that the top and bottom hBN encapsulating layers are approximately equal thicknesses, ~35 nm each[19]. The oscillating signal strength comes from background room light. We calculate the electric field experienced by IXs away from the etched graphene as $E_{hs} = \frac{1}{2} * \left(\frac{V_{tg}}{t_{t-BN}} - \frac{V_{bg}}{t_{b-BN}}\right) * \frac{\epsilon_{hBN}}{\epsilon_{hs}}$, where $E_{hs}$ is the magnitude of the electric field, $V_{t(b)g}$ is the voltage applied to the top (back) gate, $t_{t(b)-BN}$ is the thickness of the top (back) hBN encapsulating layer, $\epsilon_{hBN}$= 3.9 is the relative dielectric constant of the hBN encapsulation layer, and $\epsilon_{hs}$ = 5.3 is the dielectric constant of the heterostructure, modelling the layers as a system of three capacitors as described by *Unuchek et al.*[24]. This calculation leads to a maximum field of $E_{hs}$ = 0.125 V/nm. The total IX energy shift is 270 meV, and the total electric field difference between the first and last frame is 0.25 V/nm, leading to a measured IX dipole moment of 1.1 nm.

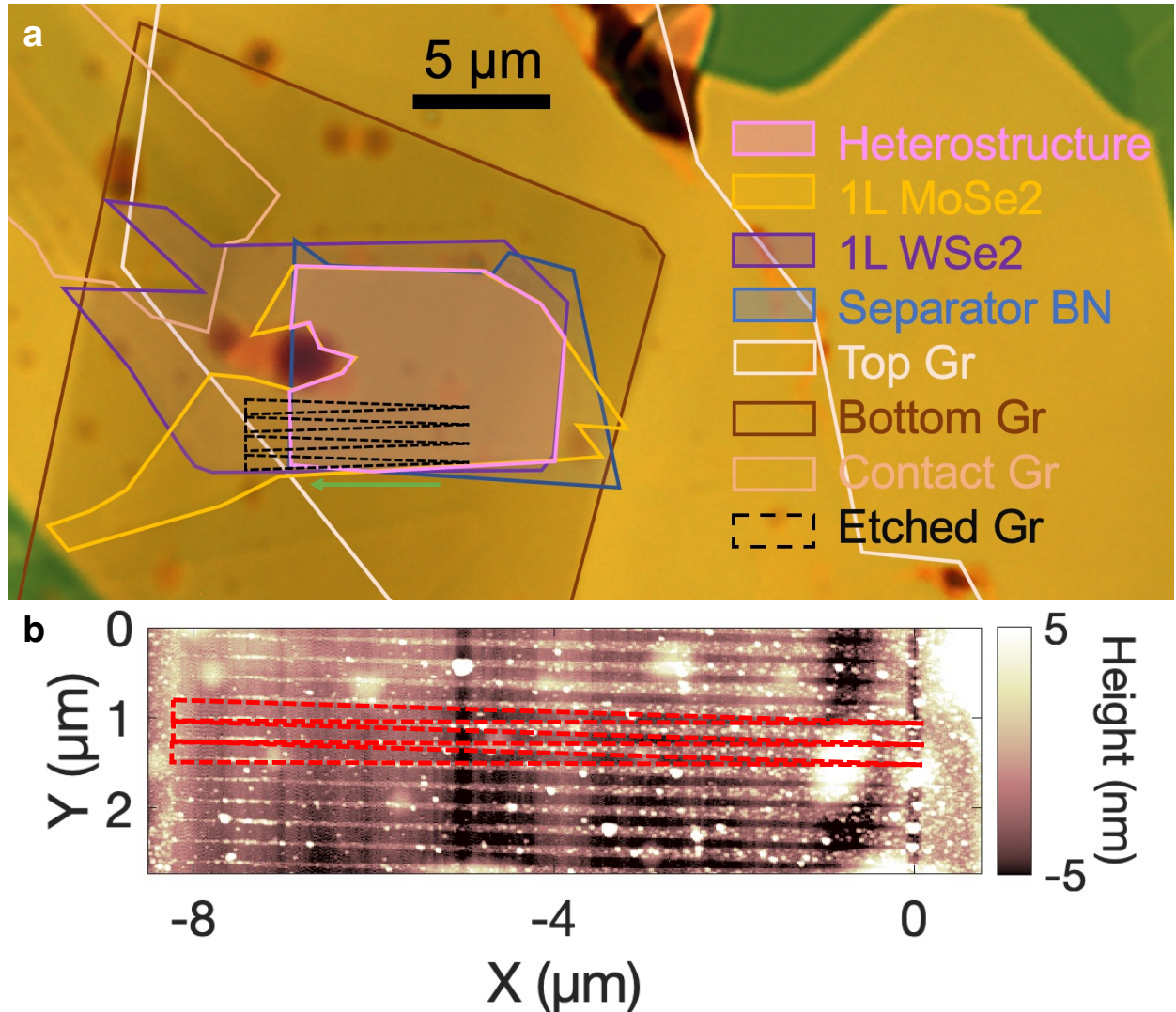

**Figure S3. Optical and AFM images of the device**. **a**, Optical microscope image of the device with layers outlined. A total of 20 slides are etched into the topgate graphene, with roughly 10 of them on the heterostructure area. Green arrow indicates direction of induced exciton current, scale bar is 5 μm. **b**, Atomic force microscopy (AFM) of the etched graphene, with some slide structures outlined in dashed red lines. Several linecuts are taken from this AFM data to measure the dimensions of the triangular etch and ensure that the etched area width increases linearly. The etched slide structures are fabricated with a length of 8.3 μm, but the hBN separator ends approximately 6 μm away from the top of the triangular etch, and the graphene etch width is 180nm at this location. The sharp tip of the triangle has a minimum width of 30 nm, which is included in the COMSOL simulation (Fig. 1c), modeling them as trapezoids instead of triangles.

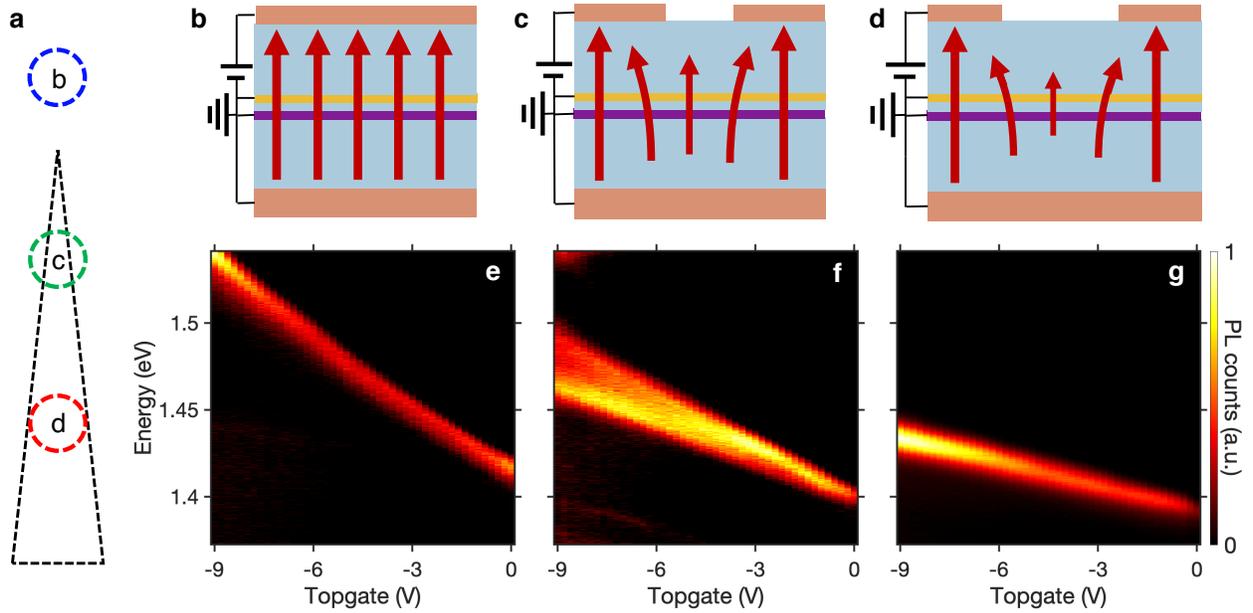

**Figure S4. Confocal PL gate maps taken off and on different parts of the slide structure. a,** Cartoon of the etched graphene (black dashes) and the different confocal PL collection spots off (b) and on (c, d) the etched graphene, with confocal collection spot in the same location as the excitation laser. **b-d,** Depictions of the electric field lines (red arrows) for no etched graphene (b), a narrow channel of etched graphene (c), and a wider channel of etched graphene (d), approximately equivalent to the PL collection locations shown in (a). **e-g,** Confocal PL taken at the three locations shown in (a) with changing topgate voltage, $V_{bg} = 0$ V, excitation power = 13 μW. Our confocal PL setup uses a 15 cm lens to reimage our PL signal from the sample onto a 50 μm pinhole, with a magnification of 33x. This leads to a collection area equivalent to a 1.5 μm diameter circle on the sample. The decreasing IX energy shift from (e) to (g) for the same applied topgate voltage shows that the field underneath the etched graphene is appropriately weaker. Additionally, we note that figures (f and g) show exclusively PL from IXs experiencing a weaker field than (e), despite the expectation that there is a saddle-like potential energy between parallel etched slides where the electric field is greater than this minimum. This shows that almost all IXs within the larger confocal collection area fall into one of the potential energy channels created by the etched graphene, which is expected given the small distance between channels and steep potential energy gradient along the sides of the channels.

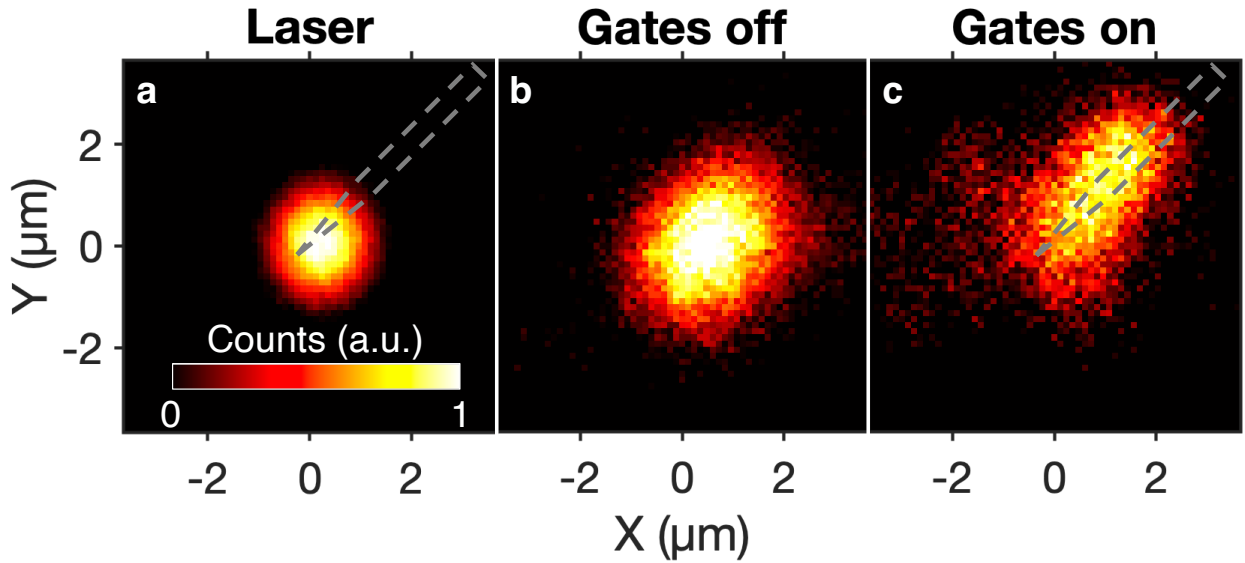

**Figure S5: Position dependent PL from a secondary slide structure.** A single nanopatterned slide of length 1.5 μm is etched on a different area of the same heterostructure. The end of the slide is connected to a nanopatterned rectangle of length 4.5 μm, which provides energetically flat 1D confinement. The nanopatterned graphene is roughly shown by the grey dashed lines. (a) shows the laser spot at the top of the slide. (b) IX PL with no applied gate voltage, showing isotropic diffusion. (c) IX PL with $V_{tg}$ = -7V, showing unidirectional drift along the patterned graphene. In this structure, we observe IX diffusion past the end of the slide into the flat channel.

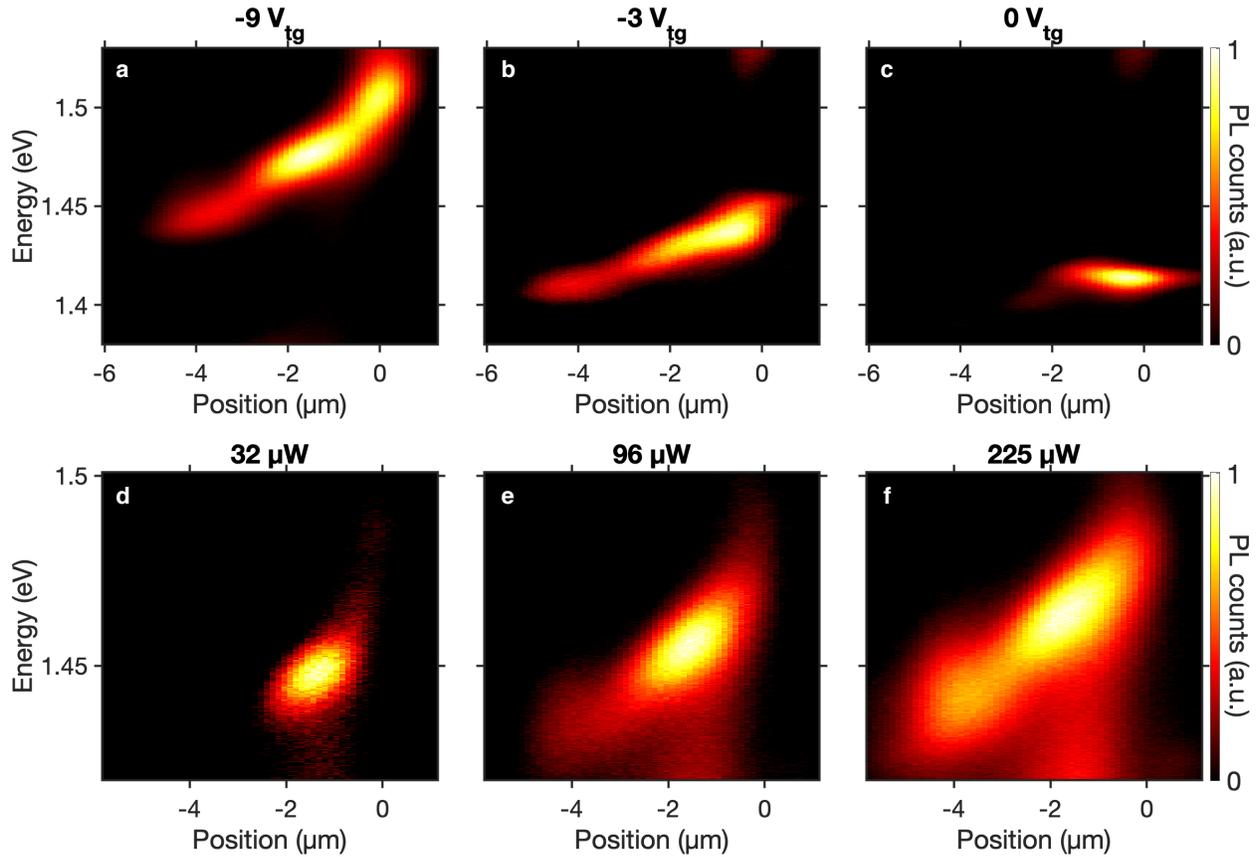

**Figure S6. Spatially and energetically resolved PL for several gate voltages and powers.**
Excitation spot is near the top of the slide structure. (a-c) have P = 225 μW, (d-f) have $V_{tg}$ = -7 V.
Spectrally resolved PL images are integrated over energy to generate the data in Figs. 2 (b, d).
Comparing (d) to (f), we note that the change in IX energy due to IX-IX repulsion is ~30 meV,
which is comparable but less than the change in IX energy from the top to bottom of the slide, ~70
meV, for the strongest measured gate voltage in (a), $V_{tg}$ = -9 V. Slide top is positioned at x = 0 μm.

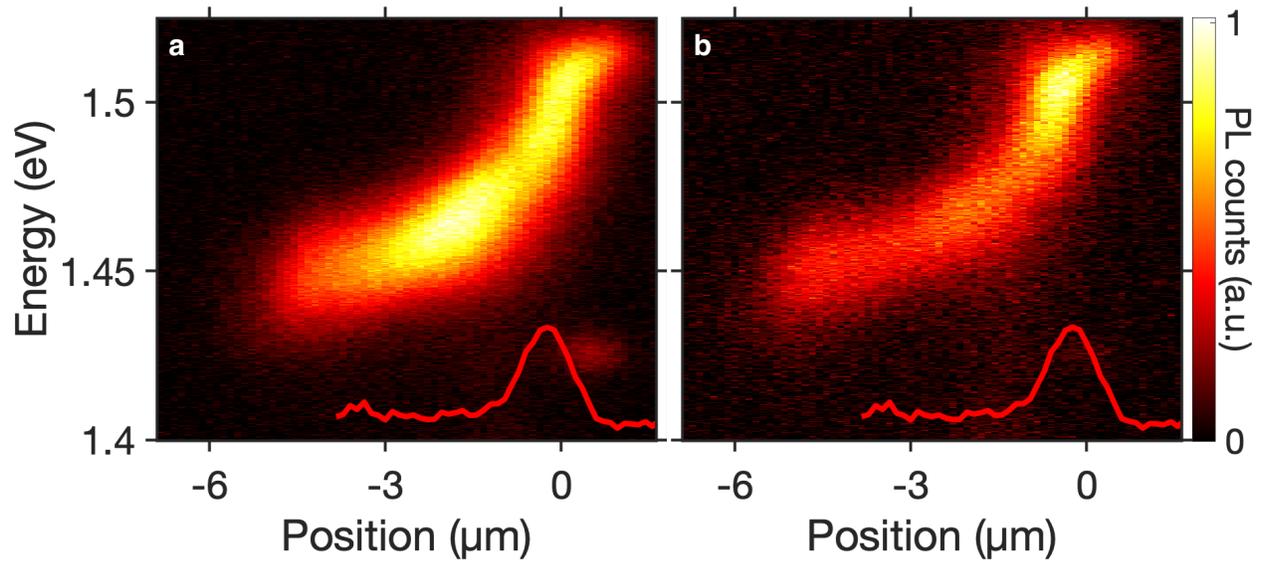

**Figure S7. Circularly polarized spectral image PL**. Co-circular (a) and cross-circular (b) polarization between excitation laser and detection. Red lines show laser excitation profile, a 720 nm, 50 µW power. These plots are integrated over energy to generate the data in Fig. 2c. Slide top is positioned at x = 0 µm.

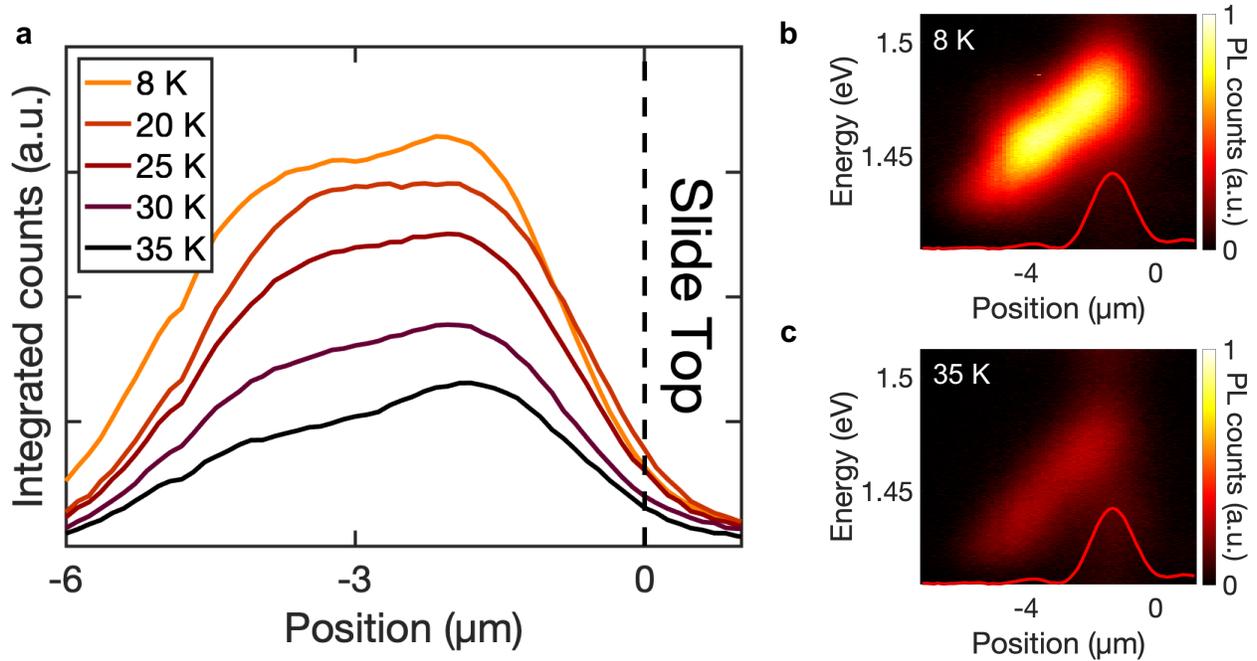

**Figure S8. Temperature dependent spatially resolved PL.** (a) Spatially integrated PL counts on the slide structure as a function of sample temperature, with $V_{tg}$ = -7V, excitation power = 350 μW at 670 nm. We observe that the spatial PL remains largely unchanged up to 20 K, and then uniformly reduces in intensity across the device up to 35 K and vanishes for higher temperatures. (b-c) Spatially and energetically resolved PL along the slide structure at 8 K (b) and 35 K (c). Red line shows laser excitation profile.

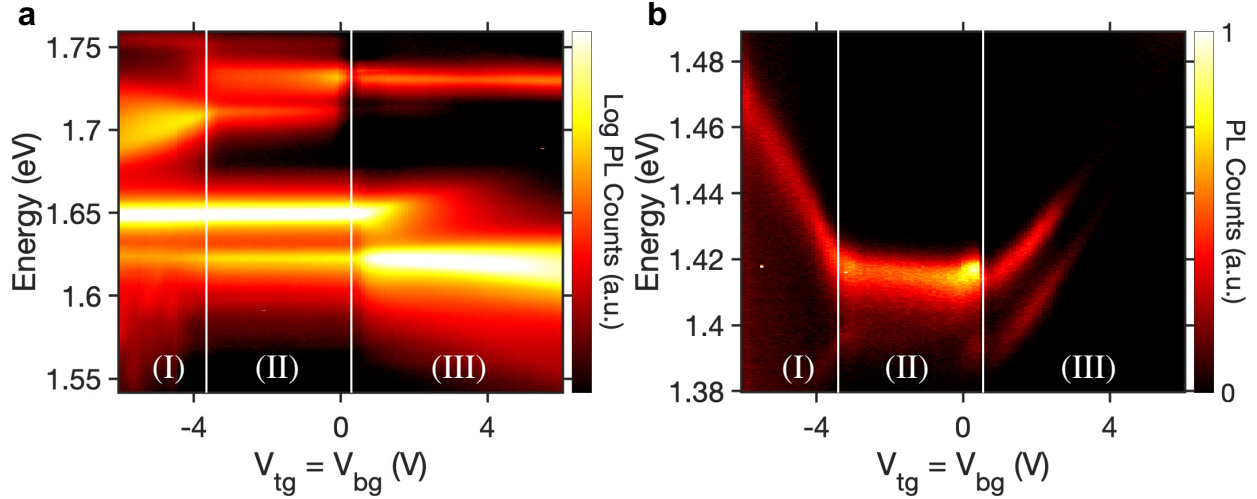

**Figure S9: Confocal doping dependent PL on the heterostructure.** (a) Intralayer exciton and (b) interlayer exciton photoluminescence signals, using equal sign and magnitude voltages applied to the top and back gates to dope the heterostructure. PL is performed on an unpatterned portion of the hBN-separated TMD heterostructure. We observe three distinct charging regimes. In region (I), we primarily observe PL from the $WSe_2$ positively charged trion (1.71 eV) and $MoSe_2$ neutral exciton (1.65 eV). In region (III), we observe intralayer PL predominantly from the $MoSe_2$ negatively charged trion (1.62 eV) and $WSe_2$ neutral exciton (1.73 eV). In region (II), we observe PL from both neutral and charged monolayer exciton states, and <5 meV shift in the interlayer exciton energy with doping.

We use a COMSOL model to estimate the spatially varying doping of the heterostructure under the slide and find that it varies proportionally to the strength of the electric field. At the top of the slide, where the electric field is strong, there is maximal doping of the heterostructure, and the sample doping reduces in magnitude along the slide towards the bottom. The maximum voltage applied in Fig. 2 of the main text is $V_{tg}$ = -7 V, $V_{bg}$ = 0 V, which corresponds to a doping level at the top of the slide nearly equivalent to $V_{tg} = V_{bg}$ = -3.5 V, which is in region (II), near the crossover to region (I). At the bottom of the slide, where the electric field is weak, the doping is close to zero, near $V_{tg} = V_{bg}$ = 0 V still in region (II), near the crossover to region (III). Thus, the doping level across the entirety of the heterostructure region in the data in the main text corresponds to the region (II) doping regime, where the sample doping accounts for only a small fraction of the shift in IX energy.

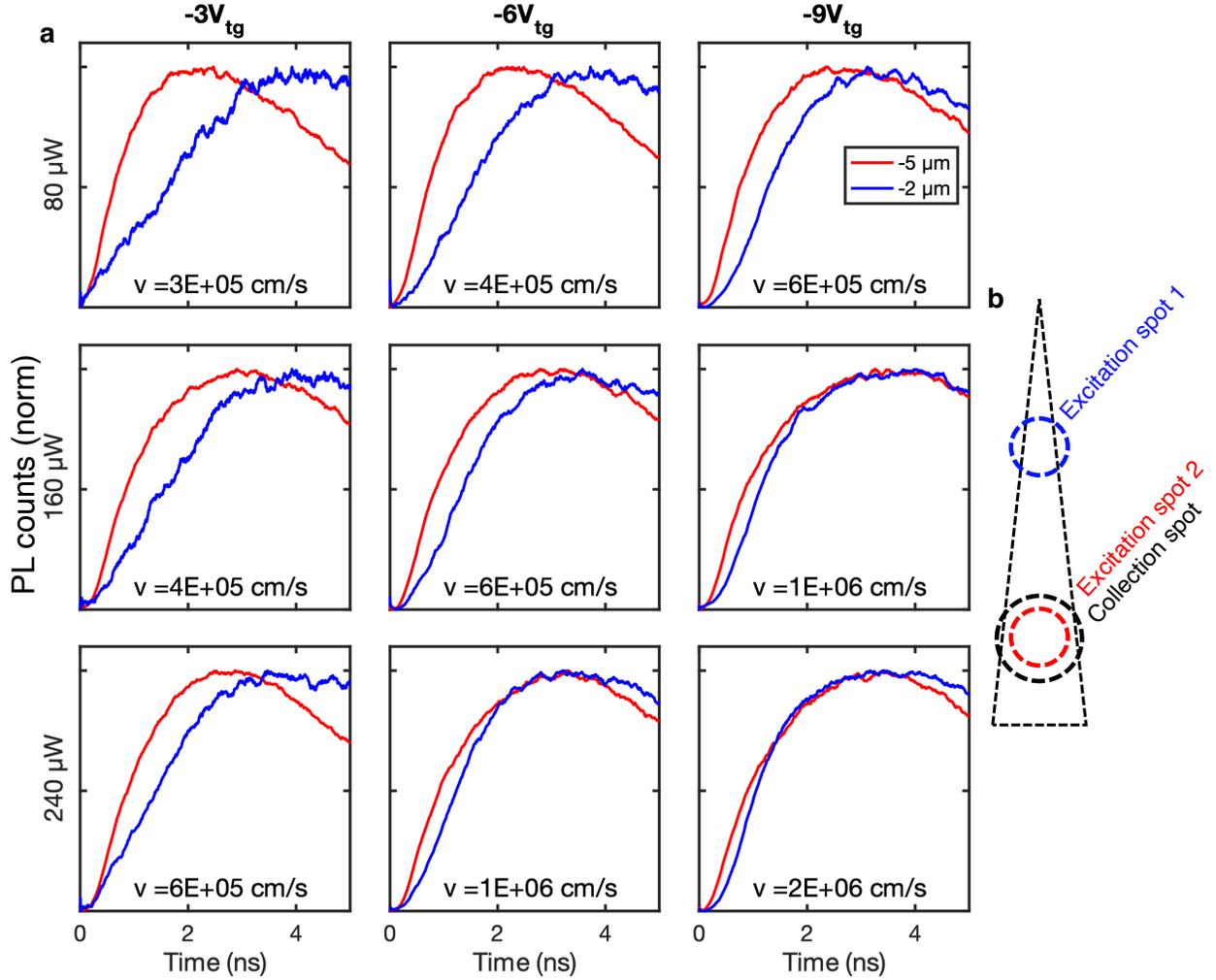

**Figure S10. Time dependent PL for several topgate voltages and powers. a**, Time resolved PL collection, showing the rise time of exciton creation and motion for excitation spots at the top and bottom of the slide. PL collection was filtered spatially and spectrally, as in main text Fig. 4a. The rows show PL with the same excitation power, listed on the left of the figure, and columns show data for the same gate voltage, listed at the top. Data is normalized to max counts of each respective scan. Legend locations refer to distance from the top of the slide. Changing the excitation power only changes the induced IX-IX repulsion force. In contrast, changing the applied gate voltage will change both the induced potential energy shift along the slide and the IX-IX repulsion, as the potential energy channel becomes deeper, pushing IX's closer together in the direction perpendicular to induced exciton flow. The velocity (v) for each scan is calculated by dividing $\Delta x$ difference in excitation position by $\Delta t$ time at which the PL reaches half of its maximum counts, as in the main text. Laser excitation was 757 nm, on the MoSe$_2$ monolayer exciton resonance, using the wavelength filtered supercontinuum laser at 78 MHz rep rate. We excite on the MoSe$_2$ resonance as we observed the rise time of IX PL taken away from the etched graphene area did not shift with gate voltage, whereas excitation at 670 or 720nm showed shift in PL rise time with changing gate voltage. **b**, Excitation and collection locations on the etched FLG.

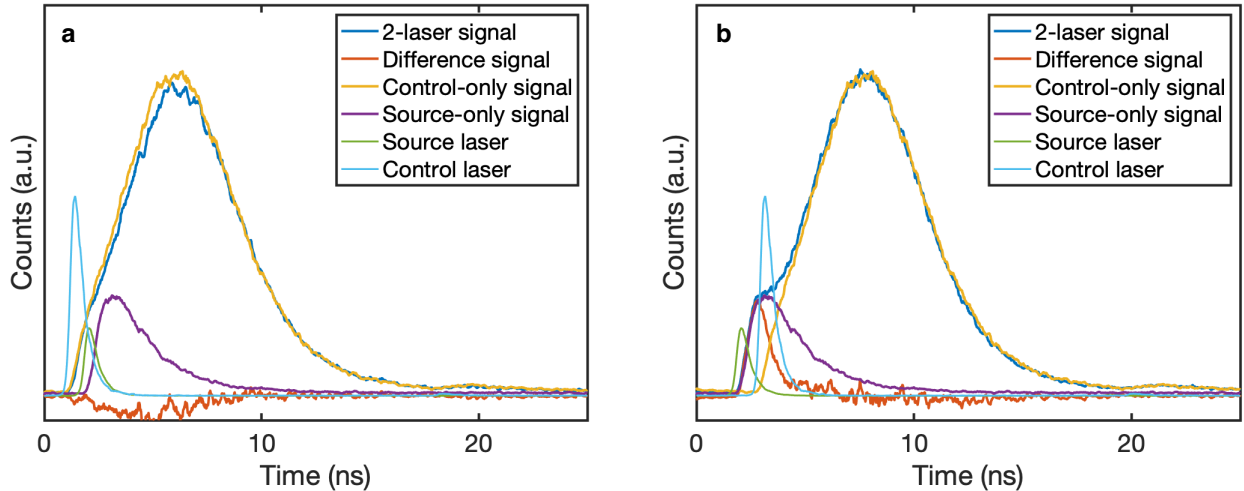

**Figure S11. Two-laser time dependent confocal PL off the slide structure.** Time resolved PL signal with higher power (80 μW) control laser before (a) and after (b) the lower power (3 μW) signal laser, to confirm that the laser power used in Fig. 4b is sufficient to saturate the PL signal within the confocal collection spot. The 80 μW, 39 MHz rep rate leads to an equivalent peak power of the 55 μW, 26 MHz rep rate control laser used in Fig 4b. Both lasers are aligned to the confocal collection spot, away from the etched graphene. The cyan (green) line shows the time delayed control (source) laser profile (laser powers not to scale). The yellow (purple) line shows the IX PL signal with only the control (source) laser on. The blue line shows the IX PL signal with both lasers on. The orange line "difference signal" is calculated by subtracting the control only (yellow) IX signal with both lasers on (blue). In (b), the source laser first excites IXs in the sample, which raises the orange difference signal to non-zero at t = 3 ns, which then returns to zero when the control laser hits the sample at t = 5 ns, saturating the IX population within the confocal collection area. In (a), the IX signal is first saturated by the control laser at t = 2 ns, and the source laser at t = 3 ns does not change the IX PL.